%% file: manuscript_mapping.tex
\documentclass[journal=jacsat,manuscript=article]{achemso}

\usepackage[version=3]{mhchem} 
\usepackage[]{mathrsfs}
\usepackage[]{listings}
\usepackage[latin1]{inputenc}
\usepackage{amssymb}
\usepackage{epsfig}
\usepackage{xspace}
\usepackage{pstricks}
\usepackage{psfrag}
\usepackage{textgreek}

\input{./reliability/symsRel}

\newcommand{\figdir}  {./figures/paper}

\author{\vspace*{-0.1cm}\it Y.Z. Lv}
\author{\vspace*{-0.1cm}\it H. Cai}
\author{\vspace*{-0.1cm}\it Y. Wu}
\author{\vspace*{-0.1cm}\it Yu.Yu. Illarionov}
\affiliation{Laboratory of 2D Optoelectronics and Nanoelectronics (L2DON), MSE Department, Southern University of Science and Technology, 1088 Xueyuan Blvd, 518055 Shenzhen, China}
\email{illarionov@sustech.edu.cn}

\title[An \textsf{achemso} demo]
  {Mapping diverse hysteresis dynamics in scaled MoS$_2$ FETs using the universal method derived from TCAD modeling}

\begin{document}



\begin{abstract}
Field-effect transistors (FETs) based on 2D materials have already reached the stage of trial FAB integration. However, reliability limitations caused by various defects present a serious obstacle for their smooth way forward, especially when scaling the device geometries. Still the ongoing research is mostly focused on pure technology aspects, while reliability is often recalled only when showing a randomly measured gate transfer curve to manifest that the hysteresis is ``negligible''. In fact the hysteresis dynamics contain unique fingerprints of various mechanisms which may coexist or cancel each other, being more complex in scaled FETs, for instance because of simultaneous interaction of defects with the channel and top gate in thin insulators. To fill this gap, here by doing TCAD modeling for nanoscale MoS$_2$/HfO$_2$ FETs we introduce the universal hysteresis mapping method which can correctly capture commonly measured diverse hysteresis dynamics such as conventional clockwise (CW) and counterclockwise (CCW) hysteresis, as well as CW/CCW switching and time separation. Next we extend this method to bias-temperature instabilities (BTI) and show a clear correlation between complex hysteresis dynamics and abnormal BTI recovery. Finally, we validate our mapping method using available experimental data for MoS$_2$ FETs and demonstrate that it provides far more accurate results than a conventional constant current extraction of the hysteresis width, being also usable if a CCW hysteresis is caused by mobile ions.       

\end{abstract}
\section{Introduction}

Recent research advances in FETs with atomically thin 2D channels~\cite{WACHTER17,LEMME22,HUANG22,TAN23,PENDURTHI24} have made it possible to start with their trial FAB integration~\cite{ASSELBERGHS20,DOROW22,CHUNG24,MORTELMANS24}, while considering these new device technologies as a potential option to continue scaling of modern electronics following the Moore's law~\cite{SCHWIERZ15,DAS21,LEMME22}. However, the way of 2D FETs to the market is delayed by severe reliability limitations caused by preexisting and process-induced defects in gate insulators~\cite{ILLARIONOV20A} which still make it hard to compete with Si technologies in stable long-term operation. The material combination MoS$_2$/high-k oxide~\cite{JAYACHANDRAN24,CHUNG24,MORTELMANS24} which is now used as the core for 2D n-FETs is unfortunately not an exception since the upper defect bands of HfO$_2$ and Al$_2$O$_3$ are rather close to the conduction band edge of MoS$_2$~\cite{ILLARIONOV20A,ILLARIONOV24A}, not to mention possible process-induced defects~\cite{ILLARIONOV24A}. Furthermore, the reliability problem significantly complicates when the device geometries are scaled down~\cite{KACZER10A,GRASSER12A,MARKOV13}. This is because in nanoscale devices the electrostatic impact of every particular defect is larger, and thus even a relatively small number of defects or few localized defect clusters still present in mature 2D FETs could result in serious perturbations of stable device operation. As the real potential of 2D FETs could be fully exploited only for sub-10$\,$nm channel lengths~\cite{IRDS21}, further progress in 2D devices requires proper understanding of new physical mechanisms of reliability limitations which would accompany aggressive scaling. This can be achieved only if certain standard metrics for the reliability are used in the future analysis, while being easily benchmarkable using universal experimental techniques.

The commonly observed hysteresis of the gate transfer characteristics can be considered as an important metric for the reliability analysis as it is directly related to the gate bias stability. Still, the absolute majority of previous studies are restricted to fragmentary observation of hysteresis at a random sweep rate and often speculate that it is ``negligible'', ``near-zero'' or ``small''~\cite{ROH16,VU18,CHO21E,VENKA24,VENKA24,FAN25,LAN25} while substituting real physical analysis by generation of random ill-defined numbers for the hysteresis width. In reality, the hysteresis in 2D FETs can have multiple contributions which may cancel each other~\cite{PROVIAS23} and become more crucial in scaled devices, thus requiring a complex analysis. Though being disregarded in most studies, the information obtained from comprehensive hysteresis measurements could be of key importance for understanding the physical mechanisms such as simultaneous interaction of charge traps with the channel and top gate which perturb stable operation of 2D FETs. Furthermore, complementing correctly extracted hysteresis data by the results measured for bias temperature instabilities (BTI)~\cite{GRASSER11A} on the same device could significantly strengthen the reliability analysis.

Here we first perform a comprehensive TCAD modeling of hysteresis dynamics in scaled MoS$_2$/HfO$_2$ FETs and explain the changes observed with varied distance of a trap distribution from the MoS$_2$/HfO$_2$ interface. Based on these findings we introduce a universal hysteresis mapping method which can capture conventional clockwise (CW) and counterclockwise (CCW) hysteresis, as well as CW/CCW switching and time separation. Next we simulate positive BTI (PBTI) for the same trap distributions and demonstrate that complex hysteresis dynamics such as switching come together with abnormal PBTI recovery, while extending our mapping method towards the BTI analysis. Finally, we apply our universal mapping method to available experimental datasets including those for imec FAB MoS$_2$ FETs with scaled top gates~\cite{ILLARIONOV24A} and verify its functionality to capture experimentally observed hysteresis dynamics also for the case when the drift of mobile ions in oxide is involved in addition to charge trapping.

\section{TCAD setup and modeling approach}
We use a well-established TCAD simulator GTS Minimos-NT~\cite{MINIMOSNT22} which describes carrier transport in the channel with the drift-diffusion model and employs the four-state non-radiative multiphonon (NMP) model~\cite{GRASSER12A,ALKAUSKAS14} for charge trapping by oxide defects. This professional device modeling tool originally developed for Si technologies has been successfully adapted to FETs with 2D channels and used in many previous works~\cite{ILLARIONOV16B,KNOBLOCH18,KNOBLOCH23}. 

\begin{figure}[!h]
\centering
\vspace{0mm}
\begin{minipage} {\textwidth} 
\hspace{0.2cm}
\centering
  \includegraphics[width=9cm]{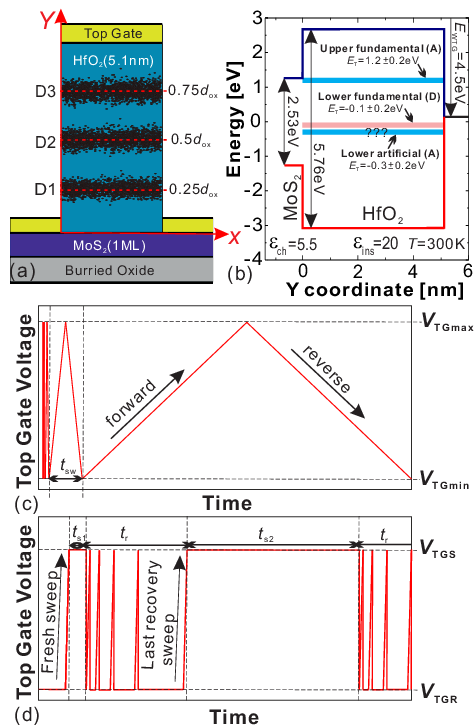} 
\caption{\label{Fig.1} (a) Schematic layout of MoS$_2$ FETs used in our TCAD simulations with trap distributions D1, D2 and D3 centered at $y_{\mathrm{T}}\,=\,0.25\,d_{\mathrm{ox}}$, 0.5$\,d_{\mathrm{ox}}$ and 0.75$\,d_{\mathrm{ox}}$ from the MoS$_2$/HfO$_2$ interface, respectively. (b) The band diagram of the channel cross-section showing the key simulation parameters in a flat band condition. We consider three defect bands: acceptor-type at $E_{\mathrm{T}}\,=\,1.2\,\pm\,$0.2$\,$eV (upper fundamental), donor-type at $E_{\mathrm{T}}\,=\,-0.1\pm\,$0.2$\,$eV (lower fundamental) and acceptor-type at $E_{\mathrm{T}}\,=\,-0.3\pm\,$0.2$\,$eV (lower artificial). (c) Schematics of top gate voltage sweeps with logarithmically increased sweep times used in our transient NMP-based simulation setup. (d) Fragmentary schematics of subsequent PBTI stress/recovery rounds with logarithmically increased stress times implemented into our TCAD tool. In reality, $\sim$10 intermediate recovery sweeps and up to 8 rounds with different $t_{\mathrm{s}}$ are used. A stabilizing stage at $V_{\mathrm{TGR}}$ is always used before the first fresh sweep. }
\end{minipage}
\end{figure}

In line with modern scaling requirements~\cite{IRDS21}, we consider top-gated single-layer MoS$_2$ FETs ($L\,=\,10\,$nm, $W\,=1\,\mu$m) with HfO$_2$ insulators of $\,d_{\mathrm{ox}}$ =5.1 nm, i.e. the equivalent oxide thickness (EOT) of 1 nm. In order to reveal the impact of trap depth on the charge trapping in a scaled oxide, we set three oxide trap distributions D1, D2, D3 of  $\sim$2000 traps centered at $y_{\mathrm{T}}\,=\,0.25\,d_{\mathrm{ox}}$, 0.5$\,d_{\mathrm{ox}}$ and 0.75$\,d_{\mathrm{ox}}$ from the MoS$_2$/HfO$_2$ interface, respectively. The schematic layout of the MoS$_2$ FETs with the positions of trap distributions in HfO$_2$ is shown in Fig.1a. The corresponding band diagram with barrier parameters taken from the literature~\cite{RASMUSSEN15} is shown in Fig.1b, with zero energy being in the middle of MoS$_2$ bandgap. It is well-known that HfO$_2$ has two fundamental defect bands~\cite{ILLARIONOV20A}, namely the upper acceptor-type (i.e. the charge -1 below $E_{\mathrm{F}}$ and 0 above $E_{\mathrm{F}}$) centered at $E_{\mathrm{T}}\,=\,1.2\,$eV and the lower donor-type (i.e. the charge +1 above $E_{\mathrm{F}}$ and 0 below $E_{\mathrm{F}}$) centered at $E_{\mathrm{T}}\,=\,-0.1\,$eV. However, since here we are targeting to get as much complex hysteresis features as possible, we also consider an artificial lower acceptor-type defect band at $E_{\mathrm{T}}\,=\,-0.3\,$eV which could express, for instance, additional process-induced defects in HfO$_2$. For all defect bands we consider the same Gaussian half-width of 0.2$\,$eV. 

The most essential ingredient of our TCAD approach is the time-dependent analysis of charge trapping which is performed using the NMP model parameters~\cite{GRASSER12A} provided in Fig.S1 in the Supplementary Information (SI). While implementing real time as a variable~\cite{KNOBLOCH18}, we first obtain the hysteresis dynamics by simulating the forward/reverse $I_{\mathrm{D}}$-$V_{\mathrm{TG}}$ curves with logarithmically spaced sweep times $t_{\mathrm{sw}}$ from 10$^{-7}\,$s to 10$^{8}\,$s, thereby going far beyond the typical measurement ranges. The schematics of these $V_{\mathrm{TG}}$ sweeps versus time is shown in Fig.1c. Next we simulate the typically used BTI measure-stress-measure loops~\cite{PROVIAS23}, as schematically shown in Fig.1d. First the device is subjected to a relaxation voltage $V_{\mathrm{TGR}}$, followed by the fast sweep of the fresh $I_{\mathrm{D}}$-$V_{\mathrm{TG}}$ curve. Then a stress voltage $V_{\mathrm{TGS}}$ is applied for a certain stress time $t_{\mathrm{s}}$, and after this the recovery is monitored using intermediate $I_{\mathrm{D}}$-$V_{\mathrm{TG}}$ sweeps while placing the device back to $V_{\mathrm{TGR}}$ in between these sweeps. For the convenience, the fresh and control sweeps are performed between $V_{\mathrm{TGR}}$ and $V_{\mathrm{TGS}}$ which also gives better convergence. Just like in real BTI experiments, subsequent stress/recovery rounds with logarithmically increased $t_{\mathrm{s}}$ were implemented in our modeling setup.  

\section{Universal mapping method: concept and verification}

In our previous studies the hysteresis width $\Delta V_{\mathrm{H}}$ was extracted using a constant current method near the threshold voltage $V_{\mathrm{th}}$ and plotted against the reciprocal sweep time 1/$t_{\mathrm{sw}}$~\cite{ILLARIONOV16A}. However, recent experimental results~\cite{PROVIAS23,KNOBLOCH23,ILLARIONOV24A} suggest that hysteresis dynamics may be more complex due to simultaneous contribution of multiple mechanisms. In line with these findings, we first consider the lower artificial defect band in HfO$_2$ and in Fig.2a-c show the $I_{\mathrm{D}}$-$V_{\mathrm{TG}}$ curves simulated for our nanoscale MoS$_2$ FETs with representative features such as pure CW hysteresis for the trap distribution D1 (Fig.2a), CW/CCW hysteresis switching for D2 (Fig.2b) and pure CCW hysteresis for D3 (Fig.2c). More $I_{\mathrm{D}}$-$V_{\mathrm{TG}}$ curves simulated for the slow and fast sweeps can be found in Fig.S2 in the SI. These results suggest that the hysteresis width $\Delta V_{\mathrm{H}}$ may depend not only on the sweep time $t_{\mathrm{sw}}$, but also on the constant current $I_{\mathrm{Dconst}}$ used for extraction. Therefore, the value of $\Delta V_{\mathrm{H}}$ obtained at a randomly selected $I_{\mathrm{Dconst}}$ may be ill-defined, for instance for the case of Fig.2b which has a switching point with zero hysteresis separating CW and CCW branches. 

\begin{figure}[!h]
\centering
\vspace{0mm}
\begin{minipage} {\textwidth} 
\hspace{0.2cm}
\centering
  \includegraphics[width=14cm]{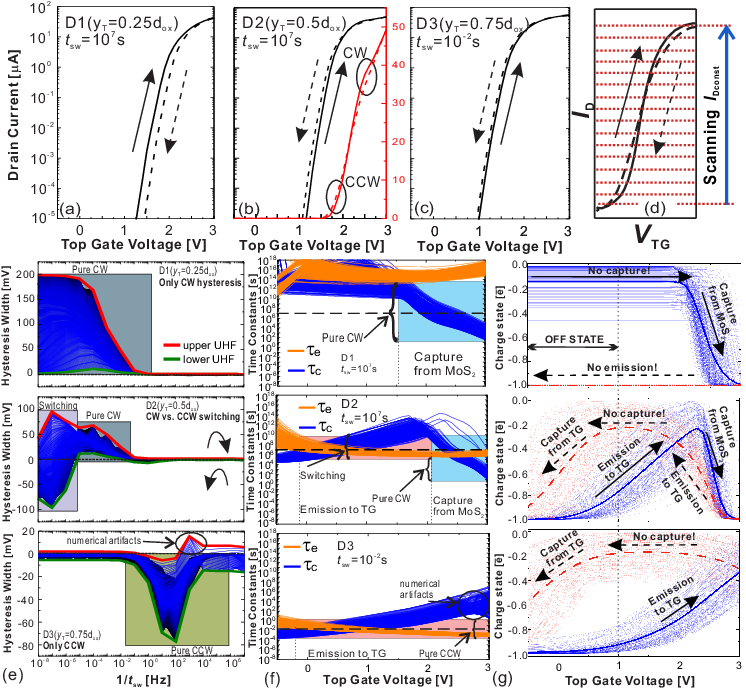} 
\caption{\label{Fig.2} The $I_{\mathrm{D}}$-$V_{\mathrm{TG}}$ curves ($V_{\mathrm{D}}\,=\,0.05\,$V, $V_{\mathrm{TG}}$ from -0.5$\,$V to 3$\,$V) simulated for MoS$_2$ FETs with lower artificial defect band in HfO$_2$ which show: (a) CW hysteresis at slow $t_{\mathrm{sw}}$ with the trap distribution D1; (b) switching between CW and CCW hysteresis at slow $t_{\mathrm{sw}}$ with the trap distribution D2; (c) CCW hysteresis at fast $t_{\mathrm{sw}}$ with the trap distribution D3. (d) Schematic illustration of our universal hysteresis mapping for the $I_{\mathrm{D}}$-$V_{\mathrm{TG}}$ curve with localized hysteresis branches of different signs. 
(e) The $\Delta V_{\mathrm{H}}$(1/$t_{\mathrm{sw}}$) dependences concluded in between upper and lower UHFs extracted from simulated $I_{\mathrm{D}}$-$V_{\mathrm{TG}}$ curves. For D1 defects only CW hysteresis is present. Mid-HfO$_2$ defects from D2 may result in a switching between CW and CCW hysteresis. D3 defects cause pure CCW hysteresis. (f) Bias dependences of the capture and emission times of the traps which are active near the hysteresis maximum as extracted from TCAD. Proximity to the top gate makes $\tau_{\mathrm{c}}$ increasing vs. $V_{\mathrm{TG}}$. (g) Variation of the charge state for these active traps during forward and reverse sweeps with the mean active charge state shown with thick lines. As it is marked in (f), those of D1 traps which are neutral can capture an electron from MoS$_2$ during forward sweep but cannot emit it during reverse sweep, thereby causing CW hysteresis. For D2 traps the CW/CCW hysteresis switching appears due to complex interaction with both MoS$_2$ and top gate. Finally, charged D3 traps emit electrons to the top gate which causes pure CCW hysteresis.}
\end{minipage}
\end{figure}

To address these limitations, here we introduce a more systematic universal hysteresis mapping method which suggests scanning of $\Delta V_{\mathrm{H}}$ from the current value slightly above $I_{\mathrm{OFF}}$ to near-$I_{\mathrm{ON}}$ values as schematically shown in Fig.2d. The exact mapping range has to be adjusted depending on the presence of obvious artifacts in a particular set of the gate transfer curves simulated or measured for different $t_{\mathrm{sw}}$, for instance noise at low currents. Using a very fine step on $I_{\mathrm{Dconst}}$, we can obtain a serie with thousands of $\Delta V_{\mathrm{H}}$(1/$t_{\mathrm{sw}}$) rather than just one. Then next we can express the hysteresis dynamics as a distribution of these curves concluded in between the upper and lower universal hysteresis functions (UHFs) constructed as a piecewise maximum and minimum from the obtained set of $\Delta V_{\mathrm{H}}$(1/$t_{\mathrm{sw}}$) dependences, respectively. Below we will show that this is the only accurate way to capture complex hysteresis dynamics such as CW/CCW switching. 

In Fig.2e we show the hysteresis mapping results for the trap distributions D1, D2 and D3 extracted from the simulated $I_{\mathrm{D}}$-$V_{\mathrm{TG}}$ curves ($t_{\mathrm{sw}}$ from 10$^{-7}$ to 10$^{8}$ s) by using 3000 $I_{\mathrm{Dconst}}$ steps. For D1 a pure clockwise hysteresis is observed for relatively slow sweeps (Fig.2e, top), which is consistent with the capture and emission time constants ($\tau_{\mathrm{c}}$ and $\tau_{\mathrm{e}}$, respectively) of active traps extracted from our TCAD simulations (Fig.2f, top). Namely, when the sweep time approaches $\tau_{\mathrm{c}}$, active traps capture an electron from the MoS$_2$ channel and change their charge state from neutral (0) to negative (-1) during the forward sweep (Fig.2e, top). Then since $\tau_{\mathrm{e}}$ is large and barely dependent on $V_{\mathrm{TG}}$, these traps remain charged during the reverse sweep. The resulting negative charge difference between forward and reverse sweeps explains the CW hysteresis which we observe for D1. 

For mid-HfO$_2$ traps in the D2 distribution, the mapping results show a CW/CCW hysteresis switching at slow $t_{\mathrm{sw}}$ (Fig.2e, center), with CW and CCW contributions being expressed by the upper and lower UHFs, respectively. The capture time $\tau_{\mathrm{c}}$ linearly increases with $V_{\mathrm{TG}}$ between -0.5$\,$V and 2$\,$V due to the impact of top gate, and then tends to decrease between 2$\,$V and 3$\,$V (Fig.2f, center) which is similar to D1 and related to the impact of MoS$_2$ channel. As a result, we can see that $\tau_{\mathrm{c}}$ and $\tau_{\mathrm{e}}$ bias dependences have two obvious intersection points which results in a more complicated charge trapping dynamics as compared to D1. Namely, active are those traps which are initially in the negative charge state (i.e. below $E_{\mathrm{F}}$). During the forward sweep as soon as $t_{\mathrm{sw}}$ becomes close to $\tau_{\mathrm{e}}$, these traps start to emit an electron to the top gate, thereby approaching neutral state (Fig.2g, center). Then because of the decay of $\tau_{\mathrm{c}}$ in the end of the forward sweep they can again capture an electron from the channel, and in the beginning of reverse sweep emit it to the top gate while remaining neutral until the OFF state in which capture of electron from the top gate becomes possible, though already with no impact on the hysteresis. As a result, the charge difference between the forward and reverse sweeps changes from positive near $V_{\mathrm{th}}$ to negative in ON state, which explains the observed CW/CCW switching of hysteresis as the same traps can simultaneously exchange charges with the MoS$_2$ channel (CW mechanism) and the top gate (CCW mechanism). However, for a $t_{\mathrm{sw}}$ range which lies below the two intersections of $\tau_{\mathrm{c}}$ and $\tau_{\mathrm{e}}$ curves but above the minimum of $\tau_{\mathrm{c}}$ at $V_{\mathrm{TG}}\,=\,3\,$V pure CW hysteresis is present just like for D1 as no interaction with the top gate is possible. 

Finally, for D3 we observe a moderately fast pure CCW hysteresis represented by lower UHF (Fig.2e, bottom). Since these traps are far from the MoS$_2$ channel, they can interact only with the top gate and thus $\tau_{\mathrm{c}}$ monotonically increases vs. $V_{\mathrm{TG}}$, with both time constants being smaller as compared to D1 and D2 cases. With the sweep times around the hysteresis maximum at $t_{\mathrm{sw}}$= 10$^{-2}\,$s which is close to $\tau_{\mathrm{e}}$, the charge state of active traps changes from negative to neutral during the forward sweep due to emission of electrons to the top gate, and then remains neutral during the reverse sweep as no capture is possible until the OFF state. As a result, positive charge difference between forward and reverse sweeps results in a CCW hysteresis. Obviously, these traps will not be active if $t_{\mathrm{sw}}$ is too large or to small, which explains why the hysteresis has a maximum which more or less matches the intersection point of time constant vs. $V_{\mathrm{TG}}$ curves. Furthermore, based on the analysis of the three cases we clearly see that intersection of rising $\tau_{\mathrm{c}}$($V_{\mathrm{TG}}$) with $\tau_{\mathrm{e}}$($V_{\mathrm{TG}}$) grounds the CCW hysteresis and intersection of decaying $\tau_{\mathrm{c}}$($V_{\mathrm{TG}}$) with $\tau_{\mathrm{e}}$($V_{\mathrm{TG}}$) is related to the CW hysteresis. We also note that when applying the suggested universal mapping method it is necessary to pay attention to possible presence of numerical artifacts when analyzing TCAD simulated data (e.g. Fig.2e, bottom), or to the noise features in the experimental data. 

\begin{figure}[!h]
\centering
\vspace{0mm}
\begin{minipage} {\textwidth} 
\hspace{0.2cm}
\centering
  \includegraphics[width=16cm]{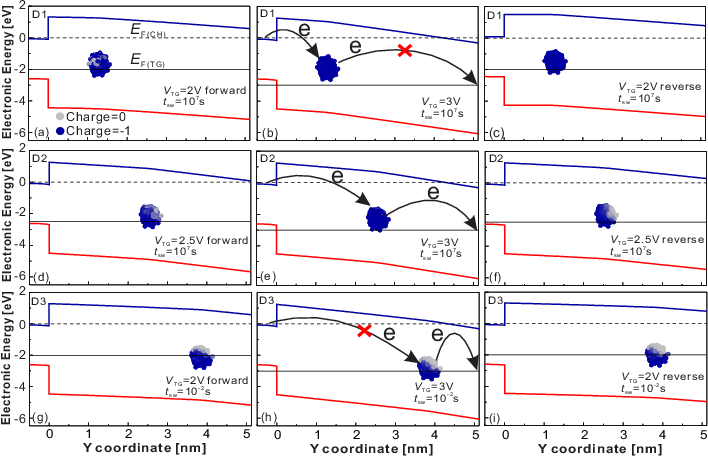} 
\caption{\label{Fig.3}The band diagrams of MoS$_2$ FETs extracted from our TCAD simulations for the forward sweep $V_{\mathrm{TG}}\,=\,2.5\,$V, $V_{\mathrm{TG}}\,=\,3\,$V and reverse sweep $V_{\mathrm{TG}}\,=\,2.5\,$V showing how the NMP trap charge state in the lower artificial defect band changes during the sweep. (a,b,c) For the distribution D1 which is closer to the MoS$_2$/HfO$_2$ interface, electron trapping from the channel results in a more negative charge state of defects during reverse sweep, which causes CW hysteresis. (d,e,f) For the mid-HfO$_2$ distribution D2, electron trapping from the channel can be followed by electron emission to the top gate, which explains the interplay between CW and CCW hysteresis. (g,h,i) The defects from distribution D3 are far from the channel, which makes electron trapping from MoS$_2$ unlikely. Thus, electron emission to the top gate makes the charge state during reverse sweep more positive and results in CCW hysteresis.}
\end{minipage}
\end{figure}

The band diagrams with all defects of the lower artificial defect band shown in Fig.3 further support our findings from Fig.2. Namely, the pure CW hysteresis is due to electron trapping from MoS$_2$ channel which saturates when all defects become negatively charged as the defect band is completely below $E_{\mathrm{F}}$ of the channel, thereby leading to a maximum at slow $t_{\mathrm{sw}}$. However, for D2 case defects can not only capture electrons from MoS$_2$ but also emit them to the top gate which has $E_{\mathrm{F}}$ crossing the lower part of the defect band which results in a CW/CCW hysteresis switching due to this complex charge trapping dynamics shown in Fig.2g, center. More band diagrams for this most complicated case can be found in Fig.S3 in the SI. For D3 case, as the traps are far from the channel, electron capture from MoS$_2$ becomes unlikely. However, since the defect band is mostly above the top gate $E_{\mathrm{F}}$ and the distance to the top gate is small, electron emission is possible which leads to the CCW hysteresis. 

\begin{figure}[!h]
\centering
\vspace{0mm}
\begin{minipage} {\textwidth} 
\hspace{0.1cm}
\centering
  \includegraphics[width=16.3cm]{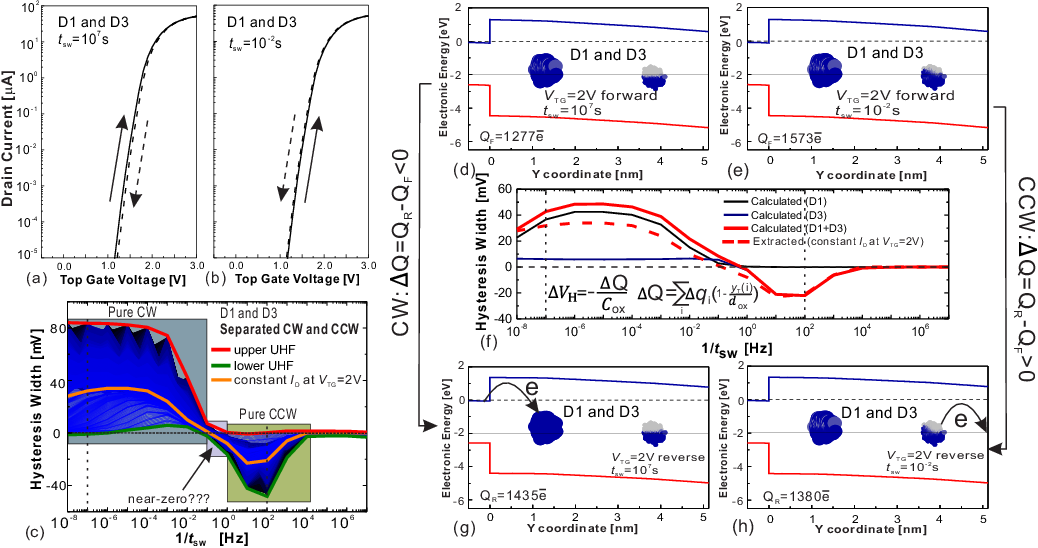} 
\caption{\label{Fig.4}$I_{\mathrm{D}}$-$V_{\mathrm{TG}}$ curves simulated for slow (a) and fast (b) sweeps assuming that both D1 and D3 are present in the oxide. (c) Mapping results show time-separated CW and CCW hysteresis on UHFs. The corresponding forward/reverse sweep band diagrams extracted from TCAD for slow (d,e) and fast (g,h) sweeps. (f) $\Delta V_{\mathrm{H}}$(1/$t_{\mathrm{sw}}$) curves calculated using the NMP charge values extracted from our TCAD simulations. D1+D3 curve is very smilar to the one obtained using a constant current method around $V_{\mathrm{TG}}\,=\,2\,$V (marked in Fig4(c)).}
\end{minipage}
\end{figure}

In Fig.4 we show that for $\sim$2000 traps equally split between D1 and D3 the CW hysteresis is present at slow sweeps (Fig.4a) and the CCW hysteresis is dominant at faster sweeps (Fig.4b). The mapping results (Fig.4c) clearly reveal time separation of these two phenomena, both having a distinct maximum in the corresponding UHF, i.e. upper for CW and lower for CCW. This case is indeed an excellent example which shows that near-zero hysteresis measured at a randomly selected single sweep time is meaningless for reliability analysis. Then we analyze the impact of NMP charges extracted from TCAD at $t_{\mathrm{sw}}$ from both peaks. Indeed, at slow $t_{\mathrm{sw}}$ the charge difference between reverse ($Q_{\mathrm{R}}$) and forward ($Q_{\mathrm{F}}$) sweeps is negative due to the dominant electron trapping from MoS$_2$ by D1 traps (Fig.4d,g). In contrast, at fast $t_{\mathrm{sw}}$ a positive $\Delta Q$ is created by dominant emission of electrons to the top gate by D3 traps (Fig.4e,h), even though their contributions to $Q_{\mathrm{R}}$ and $Q_{\mathrm{F}}$ are $\sim$3 times smaller due to (1 - $y_{\mathrm{T}}$/$d_{\mathrm{ox}}$) factor. Remarkably, the $\Delta V_{\mathrm{H}}$(1/$t_{\mathrm{sw}}$) trace calculated as -$\Delta Q$/$C_{\mathrm{ox}}$ is similar to the one extracted from the $I_{\mathrm{D}}$-$V_{\mathrm{TG}}$ curves around $V_{\mathrm{TG}}\,=\,2\,$V (Fig.4f), which justifies the validity of our approach. We note that a minor difference is because our TCAD setup considers charges in a self-consistent manner rather than a simple summing up which we perform when calculating $\Delta Q$. Finally, we have repeated these simulations assuming 450$\,$K temperature and demonstrated that both CW and CCW maxima shift towards faster sweep rates, which is consistent with the thermal activation of charge trapping~\cite{ILLARIONOV16A} (see Fig.S4 in the SI).  

\begin{figure}[!h]
\centering
\vspace{0mm}
\begin{minipage} {\textwidth} 
\hspace{0.02cm}
\centering
  \includegraphics[width=16.5cm]{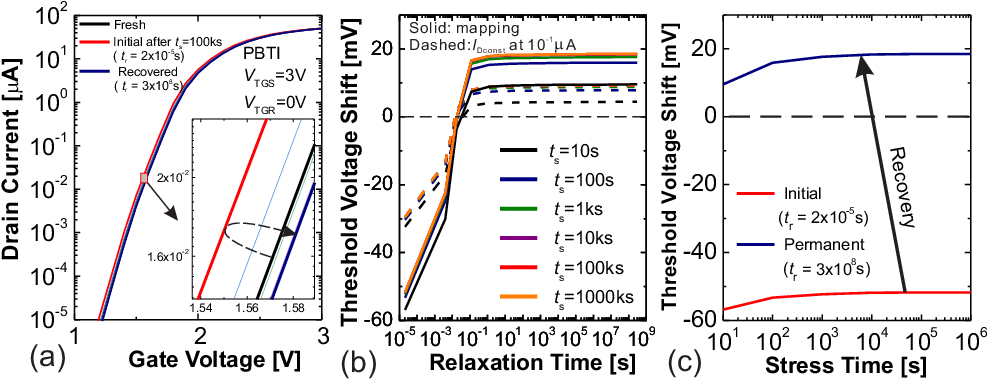} 
\caption{\label{Fig.5} (a) $I_{\mathrm{D}}$-$V_{\mathrm{TG}}$ characteristics simulated by TCAD with D1+D3 trap distributions assuming the initially stabilized condition (i.e. fresh), as well as the device after PBTI stress with $t_{\mathrm{s}}\,=\,100\,$ks, intermediate recovery points (thinner lines) and in the end of the recovery round ($t_{\mathrm{r}}\,=\,3\times 10^8\,$s). (b) The recovery traces extracted from the results simulated with six log-spaced $t_{\mathrm{s}}$ using the mapping method and constant current approach. Abnormal dynamics is due to contributions from fast charge exchange with the top gate (D3 traps) and slow charge exchange with the MoS$_2$ channel. (c) Stress time dependences of the recoverable and permanent components of PBTI as extracted from the first and last points of the traces from (b), the mapping method results are used.}
\end{minipage}
\end{figure}

In Fig.5 we show that complex hysteresis dynamics from Fig.4 is consistent with PBTI dynamics simulated using the same D1+D3 trap distributions while assuming the subsequent stress/recovery rounds (Fig.1d) with $V_{\mathrm{TGS}}\,=\,3\,$V and $V_{\mathrm{TGR}}\,=\,0\,$V. Already from the typical $I_{\mathrm{D}}$-$V_{\mathrm{TG}}$ curves (Fig.5a) of a representative stress/recovery round it is clear that PBTI drift is initially negative (i.e. NBTI-like) and becomes positive only after a long recovery. The corresponding recovery traces obtained with six log-spaced stress times $t_{\mathrm{s}}$ indeed show abnormal recovery dynamics which is not common for a usual PBTI degradation (Fig.5b). Namely, the fast component of PBTI is related to strongly recoverable negative $\Delta V_{\mathrm{th}}$. This is due to electron emission to the top gate by D3 defects and capture back during recovery, being consistent with fast CCW hysteresis in Fig.4. However, slow D1 defects simultaneously capture electrons from the MoS$_2$ channel with no emission back during recovery, and thus the permanent component is still PBTI-like which agrees with slow CW hysteresis in Fig.4. At the same time, in Fig.5c we show that both NBTI-like recoverable and PBTI-like permanent components of the degradation saturate vs. $t_{\mathrm{s}}$, which is also clear from Fig.5b. This is because for $t_{\mathrm{s}}$ above 1$\,$ks most active traps change their charge states, which is consistent with the hysteresis maxima in Fig.4c. 

We also note that just like the hysteresis width $\Delta V_{\mathrm{H}}$, the BTI shift $\Delta V_{\mathrm{th}}$ may be sensitive to the constant current point at which it is extracted. Therefore, the mapping method can be introduced for BTI in a similar way to the hysteresis, see Fig.S5 and the related discussion in the SI. While here we simulate PBTI loops with $\sim$20$\mu$s measurement delay caused by control sweeps from 0 to 3$\,$V, the recoverable component is larger if extracted closer to the subthreshold region due to increased contribution of faster traps. However, in real measurements much narrower sweep ranges just around $V_{\mathrm{th}}$ are used if the degradation is expected to be small, to reduce the delay which is still as large as hundreds of milliseconds at best. Thus we suggest that in most cases the extracted $\Delta V_{\mathrm{th}}$ should not depend on $I_{\mathrm{Dconst}}$ that much. Still in this work we show PBTI traces extracted using both approaches. For instance, it may be important to understand how the trace was extracted when trying to fit the experimental results using TCAD tools as in some cases a good fit may be obtained by lucky combination of wrong parameters.    

\section{Mapping the hysteresis and PBTI caused by fundamental HfO$_2$ defect bands}

After verifying the validity of our universal mapping method with the lower artificial defect band, we next use it for the analysis of hysteresis caused by the fundamental defect bands known for HfO$_2$~\cite{ILLARIONOV20A} while considering the same trap distributions D1, D2 and D3 (Fig.6a). The upper and lower UHFs shown in Fig.6b, top and the corresponding PBTI recovery traces for $t_{\mathrm{s}}\,=\,100\,$ks in Fig.6c, top suggest that the upper acceptor-type defect band can cause only CW hysteresis and conventional PBTI degradation. Since D1 traps are closer to the MoS$_2$/HfO$_2$ interface and thus faster, the corresponding hysteresis maximum is at faster sweep times and PBTI is more recoverable as compared to D2. We also note that the magnitude of $\Delta V_{\mathrm{th}}$ caused by PBTI is directly correlated with the fraction of hysteresis maximum which is concluded between 1/$t_{\mathrm{sw}}$ points corresponding to  $t_{\mathrm{s}}\,=\,100\,$ks and delay time $t_{\mathrm{d}}\,=\,20\,\mu$s as marked in Fig.6b. This explains why D2 traps cause larger PBTI degradation as compared to D1 traps. Furthermore, the distances between PBTI traces will start saturating vs. $t_{\mathrm{s}}$ corresponding to the position of the hysteresis maximum (e.g. Fig.5b,c). At the same time, the traps from D3 cannot interact neither with the channel being far from the MoS$_2$/HfO$_2$ interface nor with the top gate being too much above the top gate Fermi level. As a result, there is no hysteresis and PBTI induced by D3 traps. 

As for the lower donor-type defect band, we can see that being far from the conduction band of MoS$_2$, it has overall smaller contribution to both hysteresis (Fig.6b, bottom) and PBTI (Fig.6c, bottom). However, while D1 traps cause pure CW hysteresis at slow sweeps, the traps from D2 and D3 distributions can also interact with the top gate thus causing time separation of the CW/CCW hysteresis (D2) or pure CCW hysteresis (D3). Remarkably, in the latter case a purely NBTI-like PBTI degradation is observed. These findings suggest that the lower defect band may be still crucial in MoS$_2$/HfO$_2$ FETs if the defect densities are locally increased near the top gate~\cite{ILLARIONOV24A}, though being negligible as compared to the upper defect band for the channel-side defects. More detailed TCAD results for fundamental defect bands such as full hysteresis mapping with bias dependences of time constants and charged states (similar to Fig.2), band diagrams and PBTI $I_{\mathrm{D}}$-$V_{\mathrm{TG}}$ curves and traces for six $t_{\mathrm{s}}$ can be found in Fig.S6-8 and Fig.S9-11 for the upper and lower fundamental defect bands, respectively. Note that for the lower fundamental defect band interaction with the top gate is possible just like for the lower artificial band because they are both close to the top gate Fermi level, while the upper fundamental defect band is energetically far above and thus can interact only with the channel.  

\begin{figure}[!h]
\centering
\vspace{0mm}
\begin{minipage} {\textwidth} 
\hspace{0.02cm}
\centering
  \includegraphics[width=16.5cm]{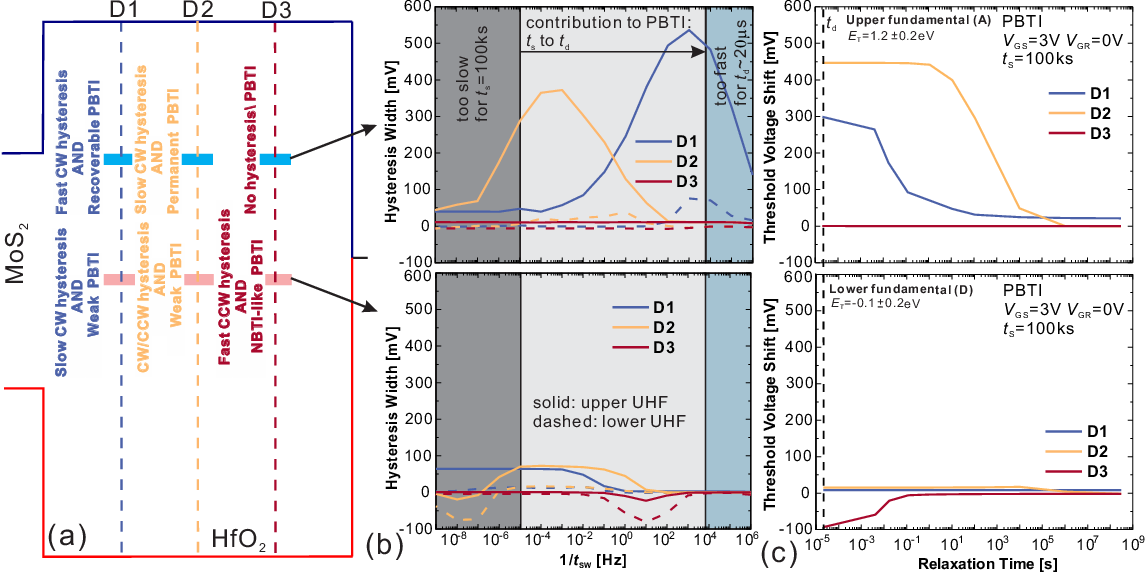} 
\caption{\label{Fig.6} (a) Schematic band diagram showing our three trap distribution within the upper and lower fundamental defect bands in HfO$_2$. The typical features of the hysteresis and BTI dynamics are noted for each case. (b) The upper and lower UHFs obtained from TCAD simulations with both defect bands. (c) The corresponding PBTI recovery traces extracted using mapping method for a representative stress time $t_{\mathrm{s}}\,=\,100\,$ks which are obviously correlated with hysteresis dynamics. The upper defect band has a considerably larger contribution because it is closer to the conduction band of MoS$_2$.}
\end{minipage}
\end{figure}

\section{Experimental validation}

We finally verify our universal hysteresis mapping method by revisiting previously published experimental results for different 2D devices. In Fig.7 we analyze the dataset obtained for the top-gated MoS$_2$/Al$_2$O$_3$ FETs in~\cite{IllARIONOV17A} which show a large CW hysteresis (Fig.7a). When applying the method to our TCAD results, we have noticed that for oxides with relatively broad defect bands the CW hysteresis width behaves as schematically shown in Fig.7b (see Fig.S12 in the SI for more details). Namely, even if a broad maximum is observed near $V_{\mathrm{th}}$, a much narrower though smaller maximum is present closer to the ON state which perfectly matches the corresponding mapping results shown in Fig.7c. There we can see that a clear maximum of $\Delta V_{\mathrm{H}}$ extracted at higher currents gets broader when moving down to $V_{\mathrm{th}}$ and finally turns to just a slight increase of the hysteresis for slow sweeps. This behaviour should be a typical feature of charge trapping in oxides with wide defect bands in which slower traps come into play during the reverse sweep, thus broadening the maximum and moving it outside the measurement range. However, previously it has been disregarded when using just a constant current extraction of $\Delta V_{\mathrm{H}}$. Thus it was typically assumed that experimental observation of the $\Delta V_{\mathrm{H}}$ maximum for 2D FETs with oxide insulators would require very long sweep times which cannot be reached in most experiments. In reality, the maximum is still there and could be used to get more information about charge trapping dynamics if the full mapping of hysteresis is performed. 

\begin{figure}[!h]
\centering
\vspace{0mm}
\begin{minipage} {\textwidth} 
\hspace{0.7cm}
\centering
  \includegraphics[width=13cm]{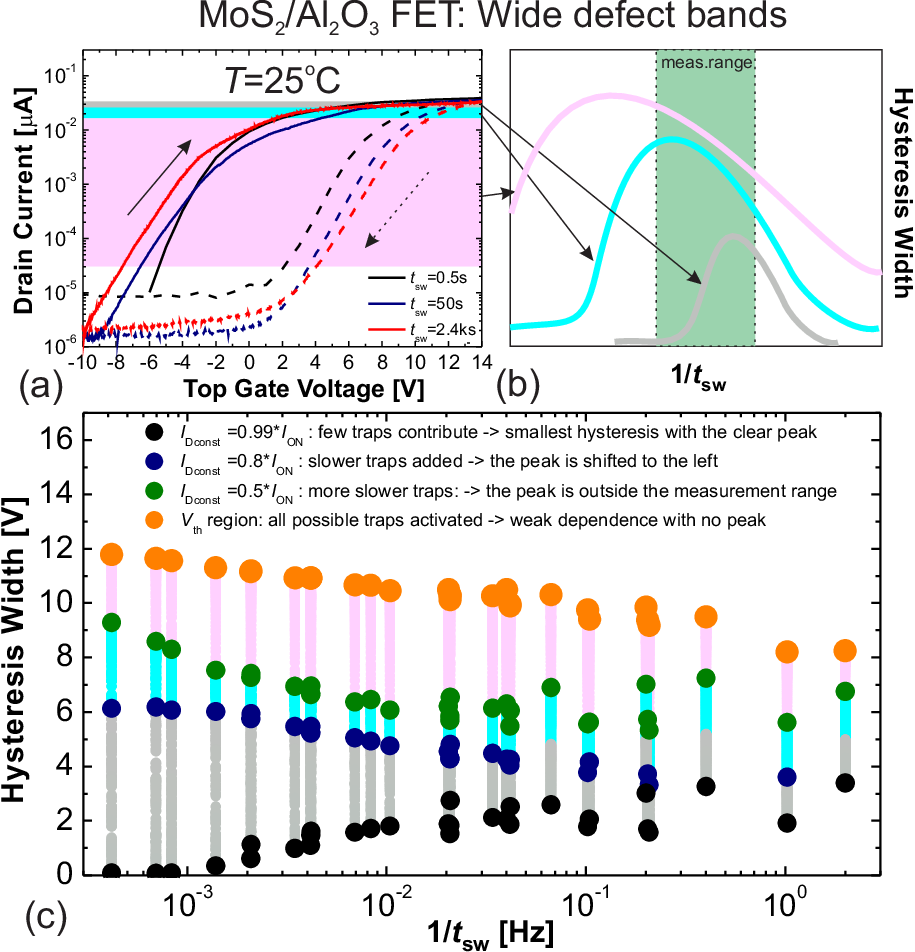} 
\caption{\label{Fig.7} New understanding of the hysteresis results obtained for the top-gated MoS$_2$/Al$_2$O$_3$ FETs in~\cite{IllARIONOV17A} using our universal hysteresis mapping method. (a) The $I_{\mathrm{D}}$-$V_{\mathrm{TG}}$ characteristics showing large CW hysteresis which actually has a maximum near the ON state as schematically shown in (b). (c) The results obtained for this dataset using our universal hysteresis mapping method confirm this schematic interpretation, i.e. a clear maximum at higher currents which becomes broader closer to $V_{\mathrm{th}}$. }
\end{minipage}
\end{figure}

Next we apply our universal hysteresis mapping method to the experimental results for imec FAB MoS$_2$ FETs~\cite{ILLARIONOV24A}, in which clockwise and counterclockwise hysteresis are present simultaneously at slow $t_{\mathrm{sw}}$ (Fig.8a), with the latter being due to top gate edge defects. A single point extraction of $\Delta V_{\mathrm{H}}$ previously allowed us to reveal this interplay only at high temperatures when it becomes more pronounced~\cite{ILLARIONOV24A}. In contrast, here we can see a sizable counterclockwise contribution in the lower UHF already at 25$^{\mathrm{o}}$C (Fig.8b), which confirms the high sensitivity of our TCAD-inspired mapping method even to those hysteresis features which are barely visible. 

Furthermore, in Fig.8c,d we examine the CW/CCW hysteresis switching which appears at high temperatures in back-gated MoS$_2$/SiO$_2$ FETs. As established in~\cite{KNOBLOCH23} by comprehensive TCAD modeling, this is due to the interplay between conventional charge trapping responsible for the CW hysteresis and drift of mobile ions in the oxide which introduce the CCW contribution. Our mapping results shown in Fig.8d clearly separate the CW contribution which appears as a standard increase of $\Delta V_{\mathrm{H}}$ for slower sweeps in the upper UHF, and the maximum of the CCW hysteresis in the lower UHF with the time position defined by the ion migration barrier~\cite{KNOBLOCH23}. The intermediate $\Delta V_{\mathrm{H}}$(1/$t_{\mathrm{sw}}$) curves such as the one extracted near the switching point are obviously affected by both hysteresis mechanisms. Therefore, the validity of our mapping method is not limited only to the hysteresis dynamics with charge trapping by insulator defects. 

\begin{figure}[!h]
\centering
\vspace{0mm}
\begin{minipage} {\textwidth} 
\hspace{0.02cm}
\centering
  \includegraphics[width=16.5cm]{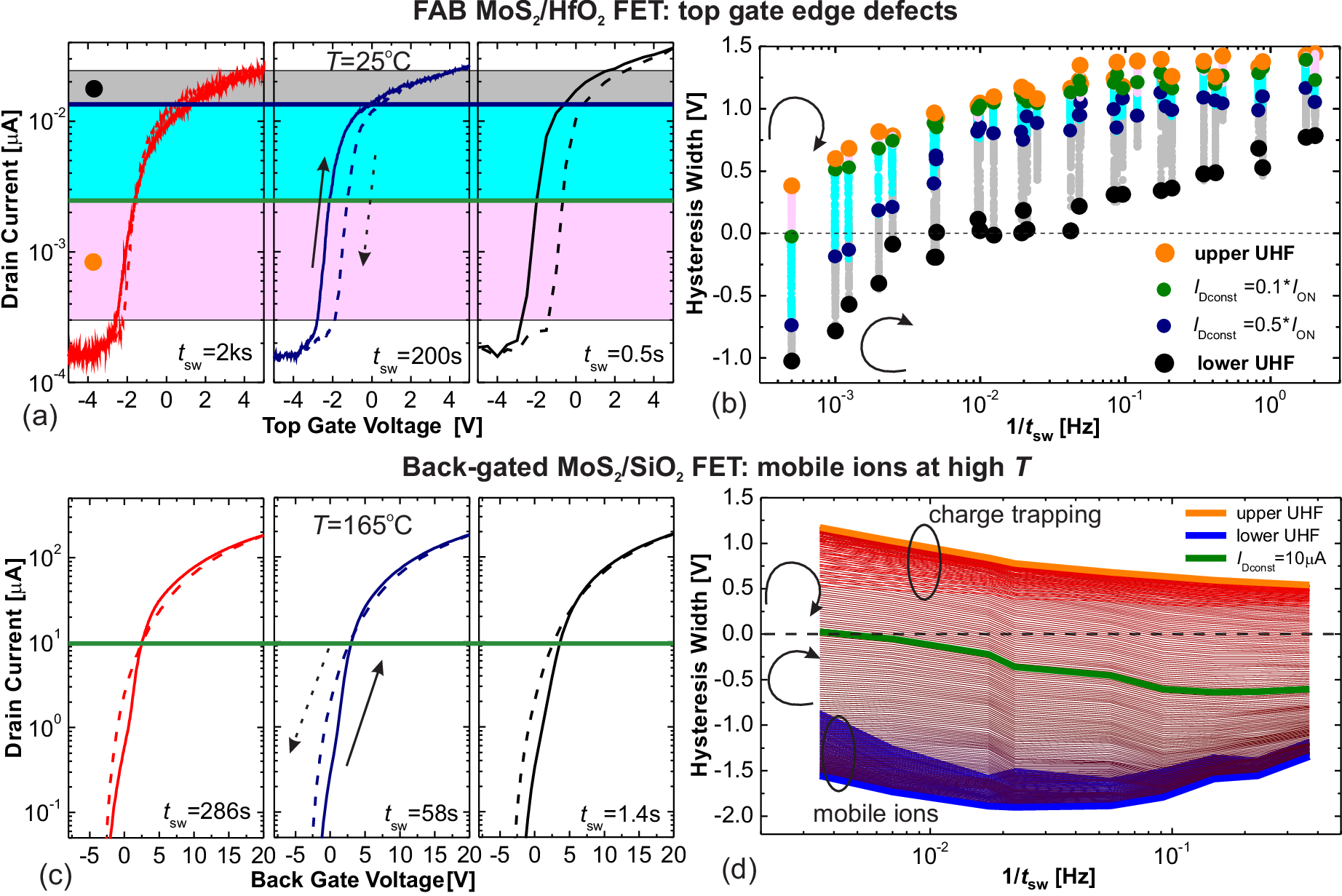} 
\caption{\label{Fig.8} (a) The $I_{\mathrm{D}}$-$V_{\mathrm{TG}}$ characteristics measured using different $t_{\mathrm{sw}}$ for FAB MoS$_2$ FETs~\cite{ILLARIONOV24A}. (b) The hysteresis curves obtained using our universal mapping method. A clear counterclockwise feature is present in the lower UHF, which reflects slow sweep hysteresis above $V_{\mathrm{th}}$. (c) The $I_{\mathrm{D}}$-$V_{\mathrm{BG}}$ characteristics measured using different $t_{\mathrm{sw}}$ for back-gated MoS$_2$/SiO$_2$ FETs~\cite{KNOBLOCH23} which show a clear CW/CCW hysteresis switching. (d) The corresponding mapping results obtained with 300 $I_{\mathrm{Dconst}}$ steps which clearly split the CW contribution coming from charge trapping by SiO$_2$ defects (upper UHF) and CCW hysteresis caused by mobile ions in SiO$_2$ (lower UHF).}
\end{minipage}
\end{figure}

\section{Conclusion}
In summary, by doing TCAD modeling for scaled MoS$_2$ FETs with different distributions of oxide traps we have derived and justified the universal method for mapping of hysteresis. The use of this method allowed us to get in depth understanding of the correlation between hysteresis and PBTI dynamics caused by HfO$_2$ defect bands while considering bias dependences of the time constants and the band diagrams. Experimental validation of our approach suggests that it can be used to explain complex hysteresis features which are observed in modern 2D FETs and often result from scaling, such as CW/CCW switching and time separation caused by an interplay between different physical mechanisms such as charge trapping from the channel and top gate, as well as drift of mobile ions in gate insulators. Our results considerably extend previous knowledge gained using the constant current extraction of hysteresis and could simplify the analysis of experimental data with TCAD, since they suggest the use of standardized hysteresis distributions rather than a single ill-defined curve. We expect that these findings will be useful to strengthen the maturity of reliability analysis in 2D FETs which are currently under trial FAB integration but cannot move forward in part due to the lack of reliability expertise in the community.

\section{Methods}

\textit{TCAD modeling approach} 

Our TCAD setup is based on a well-established simulator GTS Minimos-NT~\cite{MINIMOSNT22} which describes carrier transport in the channel with the drift-diffusion model and employs the four-state non-radiative multiphonon (NMP) model~\cite{GRASSER12A,ALKAUSKAS14} for charge trapping by oxide defects. This professional device modeling tool originally developed for Si technologies has been successfully adapted to FETs with 2D channels and used in many previous works~\cite{ILLARIONOV16B,KNOBLOCH18,KNOBLOCH23}. For a comprehensive reliability analysis it is important that Minimos-NT allows to extract the time constants and charge states for the whole ensemble of defects, as well as the band diagrams. 

We consider top-gated MoS$_2$ FETs with scaled (EOT$\,=\,$1$\,$nm) HfO$_2$ gate insulators and place the trap distributions in different sections of the oxide. Thus the most essential ingredient of our TCAD approach is the time-dependent analysis of charge trapping which is performed using the NMP model parameters~\cite{GRASSER12A} provided in Fig.S1 in the SI. While implementing real time as a variable~\cite{KNOBLOCH18}, we first obtain the hysteresis dynamics by simulating the forward/reverse $I_{\mathrm{D}}$-$V_{\mathrm{TG}}$ curves with logarithmically spaced sweep times $t_{\mathrm{sw}}$ from 10$^{-7}\,$s to 10$^{8}\,$s, thereby going far beyond the typical measurement ranges. The same approach is also employed to simulate the typically used BTI measure-stress-measure loops~\cite{PROVIAS23}, in which the device is first subjected to a relaxation voltage $V_{\mathrm{TGR}}$, followed by the fast sweep of the fresh $I_{\mathrm{D}}$-$V_{\mathrm{TG}}$ curve, and then a stress voltage $V_{\mathrm{TGS}}$ is applied for a certain stress time $t_{\mathrm{s}}$. Then the recovery is monitored using intermediate $I_{\mathrm{D}}$-$V_{\mathrm{TG}}$ sweeps while placing the device back to $V_{\mathrm{TGR}}$ in between these sweeps. Just like in real BTI experiments, subsequent stress/recovery rounds with logarithmically increased $t_{\mathrm{s}}$ were implemented in our TCAD modeling setup.  

\textit{Universal hysteresis mapping method} 

Our universal hysteresis mapping method suggests scanning of the hysteresis width  $\Delta V_{\mathrm{H}}$ from the current value slightly above $I_{\mathrm{OFF}}$ to near-$I_{\mathrm{ON}}$ values. The exact mapping range has to be adjusted depending on the presence of obvious artifacts in a particular set of the gate transfer curves simulated or measured for different $t_{\mathrm{sw}}$, for instance noise at low currents. Using a very fine step on $I_{\mathrm{Dconst}}$, we can obtain a serie with thousands of $\Delta V_{\mathrm{H}}$(1/$t_{\mathrm{sw}}$) curves. Then next we can express the hysteresis dynamics as a distribution of these curves concluded in between the upper and lower universal hysteresis functions (UHFs) constructed as a piecewise maximum and minimum from the obtained set of $\Delta V_{\mathrm{H}}$(1/$t_{\mathrm{sw}}$) dependences, respectively. 

\begin{acknowledgement}
The authors acknowledge the financial support from National Natural Science Foundation of China (NSFC) grant number W2432040, Guangdong Basic and Applied Basic Research Foundation 2024A1515010179 and Shenzhen Science and Technology Program 20231115150611001. We also thank Thomas Mueller (TU Wien), Eric Pop (Stanford), Ben Kaczer (imec) and Quentin Smets (imec) for previous collaboration which resulted in gaining the experimental datasets used for validation of the universal hysteresis mapping method.    
\end{acknowledgement}

\section{Competing interests} 
The authors declare no competing financial or non-financial interests.

\section{Data availability} 
The data that support the findings of this work are available from the corresponding author upon reasonable request.

\section{Author contributions}
Y.Z.L. performed TCAD modeling, analyzed the results and wrote the manuscript together with Y.Y.I. who also supervised the research. H.C. and Y.W. contributed to data analysis and preparation of figures.  


\providecommand{\latin}[1]{#1}
\makeatletter
\providecommand{\doi}
  {\begingroup\let\do\@makeother\dospecials
  \catcode`\{=1 \catcode`\}=2 \doi@aux}
\providecommand{\doi@aux}[1]{\endgroup\texttt{#1}}
\makeatother
\providecommand*\mcitethebibliography{\thebibliography}
\csname @ifundefined\endcsname{endmcitethebibliography}
  {\let\endmcitethebibliography\endthebibliography}{}

\end{document}


\clearpage

\section{NMP model parameters}

\begin{figure}[!h]
\hspace{0cm}
\vspace{-3mm}
\begin{minipage} {\textwidth} 
\hspace{5cm}
  \includegraphics[width=8cm]{\figdir/Fig.S1.eps} 
\caption{\label{Fig.1} A screenshot of NMP model parameters taken from our simulation script in Minimos-NT.}
\vspace*{1cm}
\end{minipage}
\end{figure}

In Fig.S1 we show a screenshot with the key parameters of the four-state NMP model~\cite{GRASSER12A} implemented into our TCAD setup. The detailed description of the meaning of these parameters can be found in the manual of Minimos-NT~\cite{MINIMOSNT22}. In this particular case, the setup with lower artificial defect band at $E_{\mathrm{T}}\,=\,-0.3\,$eV  (parameter $E_1$) is used. The Gaussian half-width of the defect band is 0.2$\,$eV (parameter $E_{\mathrm{1sigma}}$) and the type of defect band is acceptor (baseCharge\,=\,-1). 

\section {Fast and slow sweep $I_{\mathrm{D}}$-$V_{\mathrm{TG}}$ characteristics}

\begin{figure}[!h]
\hspace{0cm}
\vspace{-3mm}
\begin{minipage} {\textwidth} 
\hspace{1cm}
  \includegraphics[width=15cm]{\figdir/FigureSI1.eps} 
\caption{\label{Fig.2} Double sweep $I_{\mathrm{D}}$-$V_{\mathrm{TG}}$ characteristics of our MoS$_2$ FETs with different trap distributions simulated using $t_{\mathrm{sw}}\,=\,10^6\,$s (a-c) and $t_{\mathrm{sw}}\,=\,10^{-6}\,$s (d-f).}
\vspace*{1cm}
\end{minipage}
\end{figure}

In Fig.S2 we show the $I_{\mathrm{D}}$-$V_{\mathrm{TG}}$ characteristics simulated for our MoS$_2$ FETs using slow and fast sweep times while considering trap distribution D1, D2 and D3 and the artificial lower defect band. For the slow sweeps, the hysteresis changes from a sizable CW hysteresis for D1 (Fig.S2a) to CW/CCW switching for D2 (Fig.S2b) and then becomes negligible when the traps are closer to the top gate for D3 (Fig.S2c). For the fast sweeps (Fig.S2d-f), the hysteresis dynamics are ignorable for all three trap distributions.

\section {Band diagrams for mid-channel traps with complicated charge trapping dynamics}

\begin{figure}[!h]
\hspace{0cm}
\vspace{-3mm}
\begin{minipage} {\textwidth} 
\hspace{0.5cm}
  \includegraphics[width=15.5cm]{\figdir/FigS_D2_BD.eps} 
\caption{\label{Fig.3} Band diagrams for mid-HfO$_2$ traps from the D2 distribution which illustrate the complex hysteresis dynamics at slow sweeps caused by electron emission to the top gate and electron capture from the channel. The lower artificial defect band is considered. }
\vspace*{1cm}
\end{minipage}
\end{figure}

In Fig.S3 we show detailed band diagrams extracted from the TCAD setup with $t_{\mathrm{sw}}\,=\,10^7\,$s for the trap distribution D2 with lower artificial defect band. As illustrated in Fig.2 in the main text, during the forward sweep emission of electrons from negatively charged defects to the top gate is followed by electron capture from the MoS$_2$ channel by those defects which have become neutral. Then nearly all defects become negatively charged at the upper limit of the sweep which is $V_{\mathrm{TG}}\,=\,3\,$V. As a result, in the beginning of the reverse sweep the current in the on state is smaller than in the end of the forward sweep, and some CW hysteresis is present. Then since the emission of electrons to the top gate continues during the reverse sweep, the charge state in the reverse state becomes more positive as compared to the forward sweep (e.g. compare $V_{\mathrm{TG}}\,=\,1.5\,$V cases). As this charge state sets the $V_{\mathrm{th}}$, during the reverse sweep it is more negative and thus the CCW hysteresis is observed closer to $V_{\mathrm{th}}$. This finally brings to the CW/CCW hysteresis switching with an intersection point in which the hysteresis width is zero.  

\section {Temperature dependence of the time separated hysteresis dynamics}

In Fig.S4 we compare the dynamics of time separated hysteresis simulated using simultaneous presence of the near channel trap distribution D1 and the near top gate trap distribution D3 at $T\,=\,300\,$K and 450$\,$K. As expected~\cite{ILLARIONOV16A}, at higher tempearture the maxima of both clockwise and counterclockwise hysteresis move to the faster sweep region. This is because the time constants of oxide traps become smaller at higher temperatures, i.e. the charge trapping is thermally activated.

\begin{figure}[!h]
\hspace{0cm}
\vspace{-3mm}
\begin{minipage} {\textwidth} 
\hspace{1cm}
  \hspace{1cm} \includegraphics[width=12cm]{\figdir/Fig.S_T.eps} 
\caption{\label{Fig.4} Upper and lower UHFs extracted from TCAD results simulated assuming $T\,=\,300\,$K and 450$\,$K for the trap distributions D1+D3. Both CW and CCW hysteresis maxima are shifted to the faster sweep times. The lower artificial defect band is considered. }
\vspace*{1cm}
\end{minipage}
\end{figure}

\section {Mapping the BTI recovery dynamics}

In Fig.S5 we illustrate the BTI mapping method which can be introduced similarly to the hysteresis mapping, considering that the threshold voltage shift $\Delta V_{\mathrm{th}}$ may depend on the constant current point used for extraction. The results correspond to the D1+D3 trap distributions which cause the time separated hysteresis dynamics (e.g. Fig.S4). Then by scanning the constant current from the subthreshold region up to the ON state as shown in Fig.S5a, we can get a complicated PBTI recovery dynamics (Fig.S5b) with recoverable NBTI-like component coming from D3 and permanent PBTI-like component introduced by D1. 

\begin{figure}[!h]
\hspace{0cm}
\vspace{-3mm}
\begin{minipage} {\textwidth} 
\hspace{0.5cm}
  \includegraphics[width=15cm]{\figdir/Fig.S_BTI_mapping.eps} 
\caption{\label{Fig.5} (a) $I_{\mathrm{D}}$-$V_{\mathrm{TG}}$ characteristics simulated by TCAD assuming the initially stabilized condition (i.e. fresh), as well as the device after PBTI stress with $t_{\mathrm{s}}\,=\,100\,$ks and in the end of the recovery round ($t_{\mathrm{r}}\,=\,3\times 10^8\,$s). The trap distributions D1+D3 with lower artificial defect band are considered. Mapping of $\Delta V_{\mathrm{th}}$ is shown schematically. (b) The corresponding recovery traces extracted using the mapping method for BTI which show complicated dynamics due to contributions from fast charge exchange with the top gate (D3 traps) and slow charge exchange with the MoS$_2$ channel.}
\vspace*{1cm}
\end{minipage}
\end{figure}

Although qualitatively all PBTI recovery traces in Fig.S5b look similar, it is obvious that the exact magnitude of  $\Delta V_{\mathrm{th}}$ depends on the constant current value used for extraction. In this particular case, we use 500 mapping steps and then similarly to upper and lower UHFs for hysteresis can introduce lower and upper recovery functions, i.e. LRF and URF, respectively. Then for positive $\Delta V_{\mathrm{th}}$ it would be convenient to operate with URF and for negative $\Delta V_{\mathrm{th}}$ with LRF. However, since for BTI analysis a single recovery trace is desired for each $t_{\mathrm{s}}$, assuming possible complex recovery dynamics just like in this case, we have to introduce a combined recovery function (CRF) which is equal to LRF if $\Delta V_{\mathrm{th}}$ is negative and to URF if it is positive. As can be seen from Fig.S5b, a constant current $\Delta V_{\mathrm{th}}$ trace extracted at 0.1$\,\mu$A has a different magnitude as compared to LRF, URF and CRF, though being qualitatively similar. This has to be taken into account when trying to reproduce experimentally measured traces with TCAD as a standartized extraction method may be needed. 

\section {Detailed results for the upper fundamental defect band}

In Fig.S6 we show detailed hysteresis results for the upper fundamental defect band. From the mapping results (Fig.S6a) it is clear that D1 and D2 distributions cause the CW hystersis. The maximum of hysteresis corresponds to the time range below the intersection point of the $\tau_{\mathrm{c}}$ and $\tau_{\mathrm{e}}$ bias dependences but above the minimum of  $\tau_{\mathrm{c}}$ which is achieved at $V_{\mathrm{TG}}\,=\,3\,$V. Since this CW hysteresis comes from capture of electrons from the MoS$_2$ channel (Fig.S6c), it appears at faster sweeps for D1 traps which are closer to the MoS$_2$/HfO$_2$ interface.  As for D3 traps, they produce no hysteresis and thus only numerical noise is present on the mapping results (Fig.S6a, bottom). The corresponding $\tau_{\mathrm{c}}$ and $\tau_{\mathrm{e}}$ bias dependences (Fig.S6b, bottom) have no intersection point and thus there are no active traps, i.e. all of them remain in a neutral state during the whole sweeps (Fig.S6c, bottom). 

In Fig.S7 we show the band diagrams related to the Fig.S6. Here we see that in the D1 and D2 cases the CW hysteresis is caused by electron trapping from the channel. At the same time, for D3 there is no obvious charge trapping related to the channel or top gate, and the charge state of all traps is neutral. This is because this distribution is too far from the MoS$_2$/HfO$_2$ interface while also being completely above the top gate Fermi level.  

\begin{figure}[!h]
\hspace{0cm}
\vspace{-3mm}
\begin{minipage} {\textwidth} 
\hspace{0.2cm}
  \includegraphics[width=16cm]{\figdir/Fig_S_upper_tau.eps} 
\caption{\label{Fig.6} (a) The $\Delta V_{\mathrm{H}}$(1/$t_{\mathrm{sw}}$) dependences concluded in between upper and lower UHFs extracted from simulated $I_{\mathrm{D}}$-$V_{\mathrm{TG}}$ curves for the upper fundamental defect band. For D1 and D2 defects only CW hysteresis is present, which becomes slower in the latter case. For D3 defects no hysteresis is present but only numerical noise. (b) Bias dependences of the capture and emission times of all traps traps and those which are active near the hysteresis maximum as extracted from TCAD. (g) Variation of the charge state for active traps during forward and reverse sweeps with the mean active charge state shown with thick lines. }
\vspace*{1cm}
\end{minipage}
\end{figure}

In Fig.S8 we show the corresponding PBTI results simulated using the upper fundamental defect band while considering stress/recovery loops with six log-spaced $t_{\mathrm{s}}$. Being consistent with the CW hysteresis results, D1 and D2 distributions produce conventional PBTI recovery dynamics with considerably slower recovery of D2 traps. At the same time, PBTI degradation coming from D3 traps is negligible which is also consistent with the absence of hysteresis in that case.  

We note that if the PBTI recovery is far slower than the delay time, the difference between constant current $\Delta V_{\mathrm{th}}$ and the one extracted using the mapping method is not that large. Otherwise $\Delta V_{\mathrm{th}}$ will be more dependent on the extraction point because the recovery will take place during the control sweep. 

\begin{figure}[!h]
\hspace{0cm}
\vspace{-3mm}
\begin{minipage} {\textwidth} 
\hspace{0.2cm}
  \includegraphics[width=16cm]{\figdir/Fig.S_BD_upper.eps} 
\caption{\label{Fig.7} The band diagrams extracted from TCAD for the distributions D1(a-c), D2(d-f) and D3(g-i) considering the upper fundamental defect band and the sweep times near the hysteresis maximum. In the first two cases we see a negative charge difference between the forward and reverse sweeps, which is consistent with the CW hysteresis. For the distribution D3 the traps remain in neutral state in all cases and thus there is no charge difference between the two sweep directions.}
\vspace*{1cm}
\end{minipage}
\end{figure}

\begin{figure}[!h]
\hspace{0cm}
\vspace{-3mm}
\begin{minipage} {\textwidth} 
\hspace{1cm}
  \includegraphics[width=15cm]{\figdir/Fig.S_PBTI_upper.eps} 
\caption{\label{Fig.8} (a) $I_{\mathrm{D}}$-$V_{\mathrm{TG}}$ characteristics simulated by TCAD for D1, D2 and D3 trap distributions considering the upper fundamental defect band. Fresh and six stressed states of the device with log-spaced $t_{\mathrm{s}}$ are considered. (b) The recovery traces extracted for the corresponding stress/recovery rounds using constant current approach and the mapping method. }
\vspace*{1cm}
\end{minipage}
\end{figure}

\section {Detailed results for the lower fundamental defect band}

In Fig.S9 we show detailed hysteresis results for the lower fundamental defect band, which is of donor type. The results are generally similar to what has been discussed in Fig.2 in the main text for the lower artificial defect band, with the exception that D2 traps cause time separation of the CW/CCW hysteresis rather than switching. At the same time, for D2 and D3 distributions the key feature of interaction with the top gate is an increase of $\tau_{\mathrm{c}}$ vs. $V_{\mathrm{TG}}$, which has been also observed for the lower artificial band. 

\begin{figure}[!h]
\hspace{0cm}
\vspace{-3mm}
\begin{minipage} {\textwidth} 
\hspace{0.2cm}
  \includegraphics[width=16cm]{\figdir/Fig_S_lower_tau.eps} 
\caption{\label{Fig.9} (a,d,i) The $\Delta V_{\mathrm{H}}$(1/$t_{\mathrm{sw}}$) dependences concluded in between upper and lower UHFs extracted from simulated $I_{\mathrm{D}}$-$V_{\mathrm{TG}}$ curves for the lower fundamental defect band. For D1 defects only slow CW hysteresis is present, while the time separation of the CW and CCW hysteresis is observed for the D2 case. D3 traps result in a fast CCW hysteresis. (b,e,g,j) The corresponding bias dependences of the capture and emission times of all traps traps and those which are active near the hysteresis maximum as extracted from TCAD. (c,f,h,k) Variation of the charge state for active traps during forward and reverse sweeps with the mean active charge state shown with thick lines.}
\vspace*{1cm}
\end{minipage}
\end{figure}

In Fig.S10 we show an example of band diagrams for the lower fundamental defect band in the case of D3 traps. It is generally similar to the lower artificial defect band, i.e. we observe charge exchange with the top gate and thus pure CCW hysteresis. However, it is important to note that in the case of donor-type defect bands we are also dealing with capture/emission of electrons, i.e. emission means change of the charge state from 0 to +1 and capture from +1 to 0. 

\begin{figure}[!h]
\hspace{0cm}
\vspace{-3mm}
\begin{minipage} {\textwidth} 
\hspace{0.2cm}
  \includegraphics[width=16cm]{\figdir/Fig.S_BD_lower.eps} 
\caption{\label{Fig.10} The band diagrams extracted from TCAD for the distribution D3 considering the lower fundamental defect band and the sweep times near the hysteresis maximum. A positive charge difference between the forward and reverse sweeps is created due to emission of electrons to the top gate, which results in the CCW hysteresis. }
\vspace*{1cm}
\end{minipage}
\end{figure}

In Fig.S11 we show the corresponding PBTI results simulated using the lower fundamental defect band while considering stress/recovery loops with six log-spaced $t_{\mathrm{s}}$. Being consistent with slow CW hysteresis, the distribution D1 produces fully permanent PBTI. For D2 some decrease of $\Delta V_{\mathrm{th}}$ is observed for longer stress times because emission of electrons to the top gate comes into play, which is consistent with the CCW hysteresis maximum at slow sweeps. However, the recovery still remains PBTI-like. Finally, D3 traps cause purely NBTI-like PBTI degradation with fast recovery, which goes in line with the fast CCW hysteresis. 

\begin{figure}[!h]
\hspace{0cm}
\vspace{-3mm}
\begin{minipage} {\textwidth} 
\hspace{1cm}
  \includegraphics[width=15cm]{\figdir/Fig.S_PBTI_lower.eps} 
\caption{\label{Fig.11} (a) $I_{\mathrm{D}}$-$V_{\mathrm{TG}}$ characteristics simulated by TCAD for D1, D2 and D3 trap distributions considering the lower fundamental defect band. Fresh and six stressed states of the device with log-spaced $t_{\mathrm{s}}$ are considered. (b) The recovery traces extracted for the corresponding stress/recovery rounds using constant current approach and the mapping method. }
\vspace*{1cm}
\end{minipage}
\end{figure}

\section {Broadening of the hysteresis maximum in the mapping curves}

In Fig.S12 we consider the same D1 trap distribution with lower artificial defect band as elsewhere in the manuscript and reduce the number of $I_{\mathrm{Dconst}}$ mapping steps from 3000 to 400 to clearly illustrate the typical feature of the CW hysteresis caused by oxide defect bands which is discussed in Fig.7 in the main text. We can see that from TCAD results it also follows that a small narrow maximum of the hysteresis which could be extracted near the ON state becomes broader and shifts to slower sweep times when moving down to $V_{\mathrm{th}}$, thereby indicating that more slow traps come into play. 

\begin{figure}[!h]
\hspace{0cm}
\vspace{-3mm}
\begin{minipage} {\textwidth} 
\hspace{1cm}
  \includegraphics[width=15cm]{\figdir/Fig.S12.eps} 
\caption{\label{Fig.12} (a) Slow sweep $I_{\mathrm{D}}$-$V_{\mathrm{TG}}$ curve simulated by TCAD for D1 considering the lower artificial defect band. (b) The corresponding mapping results obtained with just 400 $I_{\mathrm{Dconst}}$ points which illustrate broadening of the hysteresis maximum closer to $V_{\mathrm{th}}$. The gray box indicates the time range inside which the observation is qualitatively similar to Fig.7 in the main text.}
\vspace*{1cm}
\end{minipage}
\end{figure}


\providecommand{\latin}[1]{#1}
\makeatletter
\providecommand{\doi}
  {\begingroup\let\do\@makeother\dospecials
  \catcode`\{=1 \catcode`\}=2 \doi@aux}
\providecommand{\doi@aux}[1]{\endgroup\texttt{#1}}
\makeatother
\providecommand*\mcitethebibliography{\thebibliography}
\csname @ifundefined\endcsname{endmcitethebibliography}
  {\let\endmcitethebibliography\endthebibliography}{}

%% file: reliability/symsRel.tex
\input{./reliability/syms2}

%% file: reliability/syms2.tex
\newcounter{myitem}

\newcommand{\resetmycnt}{\setcounter{myitem}{1}}

\resetmycnt

%% file: manuscript_mapping.bbl
\begin{mcitethebibliography}{35}
\providecommand*\natexlab[1]{#1}
\providecommand*\mciteSetBstSublistMode[1]{}
\providecommand*\mciteSetBstMaxWidthForm[2]{}
\providecommand*\mciteBstWouldAddEndPuncttrue
  {\def\EndOfBibitem{\unskip.}}
\providecommand*\mciteBstWouldAddEndPunctfalse
  {\let\EndOfBibitem\relax}
\providecommand*\mciteSetBstMidEndSepPunct[3]{}
\providecommand*\mciteSetBstSublistLabelBeginEnd[3]{}
\providecommand*\EndOfBibitem{}
\mciteSetBstSublistMode{f}
\mciteSetBstMaxWidthForm{subitem}{(\alph{mcitesubitemcount})}
\mciteSetBstSublistLabelBeginEnd
  {\mcitemaxwidthsubitemform\space}
  {\relax}
  {\relax}

\bibitem[{Wachter, S. and Polyushkin, D.K. and Bethge, O. and Mueller,
  T.}(2017)]{WACHTER17}
{Wachter, S. and Polyushkin, D.K. and Bethge, O. and Mueller, T.} {A
  Microprocessor Based on a Two-Dimensional Semiconductor}. \emph{Nat.
  Commun.} \textbf{2017}, \emph{8}, 14948\relax
\mciteBstWouldAddEndPuncttrue
\mciteSetBstMidEndSepPunct{\mcitedefaultmidpunct}
{\mcitedefaultendpunct}{\mcitedefaultseppunct}\relax
\EndOfBibitem
\bibitem[Lemme \latin{et~al.}(2022)Lemme, Akinwande, Huyghebaert, and
  Stampfer]{LEMME22}
Lemme,~M.; Akinwande,~D.; Huyghebaert,~C.; Stampfer,~C. 2D Materials for Future
  Heterogeneous Electronics. \emph{Nat. Commun.} \textbf{2022},
  \emph{13}, 1392\relax
\mciteBstWouldAddEndPuncttrue
\mciteSetBstMidEndSepPunct{\mcitedefaultmidpunct}
{\mcitedefaultendpunct}{\mcitedefaultseppunct}\relax
\EndOfBibitem
\bibitem[Huang \latin{et~al.}(2022)Huang, Wan, Shi, Zhang, Wang, Wang, Yang,
  Liu, Lin, Guan, \latin{et~al.} others]{HUANG22}
Huang,~J.-K. \latin{et~al.} {High-$\kappa$ Perovskite Membranes as
  Insulators for Two-Dimensional Transistors}. \emph{Nature} \textbf{2022},
  \emph{605}, 262--267\relax
\mciteBstWouldAddEndPuncttrue
\mciteSetBstMidEndSepPunct{\mcitedefaultmidpunct}
{\mcitedefaultendpunct}{\mcitedefaultseppunct}\relax
\EndOfBibitem
\bibitem[Tan \latin{et~al.}(2023)Tan, Yu, Tang, Gao, Yin, Zhang, Wang, Gao,
  Zhou, Zheng, Liu, Jiang, Ding, and Peng]{TAN23}
Tan,~C. \latin{et~al.} {2D Fin Field-Effect Transistors Integrated with Epitaxial High-k
  Gate Oxide}. \emph{Nature} \textbf{2023}, \emph{616}, 66--72\relax
\mciteBstWouldAddEndPuncttrue
\mciteSetBstMidEndSepPunct{\mcitedefaultmidpunct}
{\mcitedefaultendpunct}{\mcitedefaultseppunct}\relax
\EndOfBibitem
\bibitem[Pendurthi \latin{et~al.}(2024)Pendurthi, Sakib, Sadaf, Zhang, Sun,
  Chen, Jayachandran, Oberoi, Ghosh, Kumari, \latin{et~al.}
  others]{PENDURTHI24}
Pendurthi,~R.  \latin{et~al.} Monolithic
  Three-dimensional Integration of Complementary Two-Dimensional Field-Effect
  Transistors. \emph{Nat. Nanotechnol.} \textbf{2024}, \emph{19},
  970--977\relax
\mciteBstWouldAddEndPuncttrue
\mciteSetBstMidEndSepPunct{\mcitedefaultmidpunct}
{\mcitedefaultendpunct}{\mcitedefaultseppunct}\relax
\EndOfBibitem
\bibitem[Asselberghs \latin{et~al.}(2020)Asselberghs, Smets, Schram, Groven,
  Verreck, Afzalian, Arutchelvan, Gaur, Cott, Maurice, Brems, Kennes,
  Phommahaxay, Dupuy, Radisic, de~Marneffe, Thiam, Li, Devriendt, Huyghebaert,
  lin, Caymax, Morin, and Radu]{ASSELBERGHS20}
Asselberghs,~I. \latin{et~al.}  Wafer-Scale Integration of Double Gated WS$_2$
  Transistors in 300mm Si CMOS Fab. IEEE International Electron Devices Meeting
  (IEDM). 2020; pp 40--2\relax
\mciteBstWouldAddEndPuncttrue
\mciteSetBstMidEndSepPunct{\mcitedefaultmidpunct}
{\mcitedefaultendpunct}{\mcitedefaultseppunct}\relax
\EndOfBibitem
\bibitem[Dorow \latin{et~al.}(2022)Dorow, Penumatcha, Kitamura, Rogan,
  OBrien, Lee, Ramamurthy, Cheng, Maxey, Zhong, \latin{et~al.}
  others]{DOROW22}
Dorow,~C. \latin{et~al.} Gate Length
  Scaling beyond Si: Mono-Layer 2D Channel FETs Robust to Short Channel
  Effects. IEEE International Electron Devices Meeting (IEDM). 2022; pp 7--5\relax
\mciteBstWouldAddEndPuncttrue
\mciteSetBstMidEndSepPunct{\mcitedefaultmidpunct}
{\mcitedefaultendpunct}{\mcitedefaultseppunct}\relax
\EndOfBibitem
\bibitem[Chung \latin{et~al.}(2024)Chung, Chou, Yun, Hsu, Yu, Chang, Lee, Chou,
  Ho, Wei, \latin{et~al.} others]{CHUNG24}
Chung,~Y.-Y. \latin{et~al.} Stacked Channel
  Transistors with 2D Materials: An Integration Perspective. IEEE International
  Electron Devices Meeting (IEDM). 2024; pp 1--4\relax
\mciteBstWouldAddEndPuncttrue
\mciteSetBstMidEndSepPunct{\mcitedefaultmidpunct}
{\mcitedefaultendpunct}{\mcitedefaultseppunct}\relax
\EndOfBibitem
\bibitem[Mortelmans \latin{et~al.}(2024)Mortelmans, Buragohain, Kitamura,
  Dorow, Rogan, Siddiqui, Ramamurthy, Lux, Zhong, Harlson, \latin{et~al.}
  others]{MORTELMANS24}
Mortelmans,~W. \latin{et~al.} Gate
  Oxide Module Development for Scaled GAA 2D FETs Enabling SS$<$ 75mV/d and
  Record Idmax$>$ 900$\mu$A/$\mu$m at Lg$<$ 50nm. IEEE International Electron
  Devices Meeting (IEDM). 2024; pp 1--4\relax
\mciteBstWouldAddEndPuncttrue
\mciteSetBstMidEndSepPunct{\mcitedefaultmidpunct}
{\mcitedefaultendpunct}{\mcitedefaultseppunct}\relax
\EndOfBibitem
\bibitem[Schwierz \latin{et~al.}(2015)Schwierz, Pezoldt, and
  Granzner]{SCHWIERZ15}
Schwierz,~F.; Pezoldt,~J.; Granzner,~R. {Two-Dimensional Materials and Their
  Prospects in Transistor Electronics}. \emph{Nanoscale} \textbf{2015},
  \emph{7}, 8261--8283\relax
\mciteBstWouldAddEndPuncttrue
\mciteSetBstMidEndSepPunct{\mcitedefaultmidpunct}
{\mcitedefaultendpunct}{\mcitedefaultseppunct}\relax
\EndOfBibitem
\bibitem[Das \latin{et~al.}(2021)Das, Sebastian, Pop, McClellan, Franklin,
  Grasser, Knobloch, Illarionov, Penumatcha, Appenzeller, Cheng, Zhu,
  Asselberghs, Li, Avci, Bhat, Anthopoulos, and Singh]{DAS21}
Das,~S. \latin{et~al.}  {Transistors Based on Two-Dimensional Materials for
  Future Integrated Circuits}. \emph{Nat. Electron.} \textbf{2021},
  \emph{4}, 786--799\relax
\mciteBstWouldAddEndPuncttrue
\mciteSetBstMidEndSepPunct{\mcitedefaultmidpunct}
{\mcitedefaultendpunct}{\mcitedefaultseppunct}\relax
\EndOfBibitem
\bibitem[Illarionov \latin{et~al.}(2020)Illarionov, Knobloch, Jech, Lanza,
  Akinwande, Vexler, Mueller, Lemme, Fiori, Schwierz, and
  Grasser]{ILLARIONOV20A}
Illarionov,~Y. \latin{et~al.} {Insulators for
  2D nanoelectronics: the gap to bridge}. \emph{Nat. Commun.}
  \textbf{2020}, \emph{11}, 3385\relax
\mciteBstWouldAddEndPuncttrue
\mciteSetBstMidEndSepPunct{\mcitedefaultmidpunct}
{\mcitedefaultendpunct}{\mcitedefaultseppunct}\relax
\EndOfBibitem
\bibitem[Jayachandran \latin{et~al.}(2024)Jayachandran, Pendurthi, Sadaf,
  Sakib, Pannone, Chen, Han, Trainor, Kumari, Mc~Knight, \latin{et~al.}
  others]{JAYACHANDRAN24}
Jayachandran,~D. \latin{et~al.} Three-dimensional integration of two-dimensional field-effect transistors.
  \emph{Nature} \textbf{2024}, \emph{625}, 276--281\relax
\mciteBstWouldAddEndPuncttrue
\mciteSetBstMidEndSepPunct{\mcitedefaultmidpunct}
{\mcitedefaultendpunct}{\mcitedefaultseppunct}\relax
\EndOfBibitem
\bibitem[Illarionov \latin{et~al.}(2024)Illarionov, Knobloch, Uzlu,
  Banshchikov, Ivanov, Sverdlov, Otto, Stoll, Vexler, Waltl, Wang, Manna,
  Neumaier, M.C., N.S., and Grasser]{ILLARIONOV24A}
Illarionov,~Y. \latin{et~al.}  {Variability and high temperature reliability of
  graphene field-effect transistors with thin epitaxial CaF$_2$ insulators}.
  \emph{npj 2D Mater. and Appl.} \textbf{2024}, \emph{8}, 23\relax
\mciteBstWouldAddEndPuncttrue
\mciteSetBstMidEndSepPunct{\mcitedefaultmidpunct}
{\mcitedefaultendpunct}{\mcitedefaultseppunct}\relax
\EndOfBibitem
\bibitem[Kaczer \latin{et~al.}(2010)Kaczer, Roussel, Grasser, and
  Groeseneken]{KACZER10A}
Kaczer,~B.; Roussel,~P.; Grasser,~T.; Groeseneken,~G. {Statistics of Multiple
  Trapped Charges in the Gate Oxide of Deeply Scaled MOSFET Devices-Application
  to NBTI}. \emph{IEEE Electron Dev. Lett.} \textbf{2010}, \emph{31},
  411--413\relax
\mciteBstWouldAddEndPuncttrue
\mciteSetBstMidEndSepPunct{\mcitedefaultmidpunct}
{\mcitedefaultendpunct}{\mcitedefaultseppunct}\relax
\EndOfBibitem
\bibitem[Grasser(2012)]{GRASSER12A}
Grasser,~T. {Stochastic Charge Trapping in Oxides: From Random Telegraph Noise
  to Bias Temperature Instabilities}. \emph{Microelectron. Reliab.}
  \textbf{2012}, \emph{52}, 39--70\relax
\mciteBstWouldAddEndPuncttrue
\mciteSetBstMidEndSepPunct{\mcitedefaultmidpunct}
{\mcitedefaultendpunct}{\mcitedefaultseppunct}\relax
\EndOfBibitem
\bibitem[Markov \latin{et~al.}(2013)Markov, Amoroso, Gerrer, Adamu-Lema, and
  Asenov]{MARKOV13}
Markov,~S.; Amoroso,~S.; Gerrer,~L.; Adamu-Lema,~F.; Asenov,~A. {Statistical
  Interactions of Multiple Oxide Traps under BTI Stress of Nanoscale MOSFETs}.
  \emph{IEEE Electron Dev. Lett.} \textbf{2013}, \emph{34}, 686--688\relax
\mciteBstWouldAddEndPuncttrue
\mciteSetBstMidEndSepPunct{\mcitedefaultmidpunct}
{\mcitedefaultendpunct}{\mcitedefaultseppunct}\relax
\EndOfBibitem
\bibitem[IEEE(2021)]{IRDS21}
IEEE IRDS More Moore. 2021\relax
\mciteBstWouldAddEndPuncttrue
\mciteSetBstMidEndSepPunct{\mcitedefaultmidpunct}
{\mcitedefaultendpunct}{\mcitedefaultseppunct}\relax
\EndOfBibitem
\bibitem[Roh \latin{et~al.}(2016)Roh, Lee, Jin, and Lee]{ROH16}
Roh,~J.; Lee,~J.-H.; Jin,~S.~H.; Lee,~C. Negligible Hysteresis of Molybdenum
  Disulfide Field-Effect Transistors through Thermal Annealing. \emph{Journal
  of Information Display} \textbf{2016}, \emph{17}, 103--108\relax
\mciteBstWouldAddEndPuncttrue
\mciteSetBstMidEndSepPunct{\mcitedefaultmidpunct}
{\mcitedefaultendpunct}{\mcitedefaultseppunct}\relax
\EndOfBibitem
\bibitem[Vu \latin{et~al.}(2018)Vu, Fan, Lee, Joo, Yu, and Lee]{VU18}
Vu,~Q.; Fan,~S.; Lee,~S.; Joo,~M.-K.; Yu,~W.; Lee,~Y. {Near-Zero Hysteresis and
  Near-Ideal Subthreshold Swing in h-BN Encapsulated Single-Layer MoS$_2$
  Field-Effect Transistors}. \emph{2D Mater.} \textbf{2018}, \emph{5},
  031001\relax
\mciteBstWouldAddEndPuncttrue
\mciteSetBstMidEndSepPunct{\mcitedefaultmidpunct}
{\mcitedefaultendpunct}{\mcitedefaultseppunct}\relax
\EndOfBibitem
\bibitem[Cho \latin{et~al.}(2021)Cho, Pujar, Choi, Kang, Hong, Park, Baek, Kim,
  Lee, and Kim]{CHO21E}
Cho,~H.~W.; Pujar,~P.; Choi,~M.; Kang,~S.; Hong,~S.; Park,~J.; Baek,~S.;
  Kim,~Y.; Lee,~J.; Kim,~S. Direct Growth of Orthorhombic Hf$_{0.5}$Zr$_{0.5}$O$_2$ Thin
  Films for Hysteresis-free MoS$_2$ Negative Capacitance Field-Effect Transistors.
  \emph{npj 2D Mater. and Appl.} \textbf{2021}, \emph{5}, 46\relax
\mciteBstWouldAddEndPuncttrue
\mciteSetBstMidEndSepPunct{\mcitedefaultmidpunct}
{\mcitedefaultendpunct}{\mcitedefaultseppunct}\relax
\EndOfBibitem
\bibitem[Venkatakrishnarao \latin{et~al.}(2024)Venkatakrishnarao, Mishra, Tarn,
  Bosman, Lee, Das, Mukherjee, Talha-Dean, Zhang, Teo, \latin{et~al.}
  others]{VENKA24}
Venkatakrishnarao,~D. \latin{et~al.} Liquid Metal
  Oxide-Assisted Integration of High-k Dielectrics and Metal Contacts for
  Two-Dimensional Electronics. \emph{ACS Nano} \textbf{2024}, \emph{18},
  26911--26919\relax
\mciteBstWouldAddEndPuncttrue
\mciteSetBstMidEndSepPunct{\mcitedefaultmidpunct}
{\mcitedefaultendpunct}{\mcitedefaultseppunct}\relax
\EndOfBibitem
\bibitem[Fan \latin{et~al.}(2025)Fan, Yi, Deng, Zhou, Zhang, Yu, Li, Li, Wu,
  Zhou, \latin{et~al.} others]{FAN25}
Fan,~X. \latin{et~al.} 2D Edge-seeded Heteroepitaxy of Ultrathin High-$\kappa$
  Dielectric CaNb$_2$O$_6$ for 2D Field-Effect Transistors. \emph{Nat.
  Commun.} \textbf{2025}, \emph{16}, 2585\relax
\mciteBstWouldAddEndPuncttrue
\mciteSetBstMidEndSepPunct{\mcitedefaultmidpunct}
{\mcitedefaultendpunct}{\mcitedefaultseppunct}\relax
\EndOfBibitem
\bibitem[Lan \latin{et~al.}(2025)Lan, Yang, Kantre, Cott, Tripathi,
  Appenzeller, and Chen]{LAN25}
Lan,~H.-Y.; Yang,~S.-H.; Kantre,~K.-A.; Cott,~D.; Tripathi,~R.;
  Appenzeller,~J.; Chen,~Z. Reliability of high-performance monolayer MoS$_2$
  transistors on scaled high-$\kappa$ HfO$_2$. \emph{npj 2D Mater. and
  Appl.} \textbf{2025}, \emph{9}, 5\relax
\mciteBstWouldAddEndPuncttrue
\mciteSetBstMidEndSepPunct{\mcitedefaultmidpunct}
{\mcitedefaultendpunct}{\mcitedefaultseppunct}\relax
\EndOfBibitem
\bibitem[Provias \latin{et~al.}(2023)Provias, Knobloch, Kitamura, OBrien,
  Dorow, Waldhoer, Stampfer, Penumatcha, Lee, Ramamurthy, \latin{et~al.}
  others]{PROVIAS23}
Provias,~A. \latin{et~al.} Reliability Assessment of Double-Gated Wafer-Scale MoS$_2$ Field Effect
  Transistors through Hysteresis and Bias Temperature Instability Analyses.
  IEEE International Electron Devices Meeting (IEDM). 2023; pp 1--4\relax
\mciteBstWouldAddEndPuncttrue
\mciteSetBstMidEndSepPunct{\mcitedefaultmidpunct}
{\mcitedefaultendpunct}{\mcitedefaultseppunct}\relax
\EndOfBibitem
\bibitem[Grasser \latin{et~al.}(2011)Grasser, Kaczer, G{\"o}s, Reisinger,
  Aichinger, Hehenberger, Wagner, Franco, Toledano-Luque, and
  Nelhiebel]{GRASSER11A}
Grasser,~T.; Kaczer,~B.; G{\"o}s,~W.; Reisinger,~H.; Aichinger,~T.;
  Hehenberger,~P.; Wagner,~P.-J.; Franco,~J.; Toledano-Luque,~M.; Nelhiebel,~M.
  {The Paradigm Shift in Understanding the Bias Temperature Instability: From
  Reaction-Diffusion to Switching Oxide Traps}. \emph{IEEE Trans. Electron
  Devices} \textbf{2011}, \emph{58}, 3652--3666\relax
\mciteBstWouldAddEndPuncttrue
\mciteSetBstMidEndSepPunct{\mcitedefaultmidpunct}
{\mcitedefaultendpunct}{\mcitedefaultseppunct}\relax
\EndOfBibitem
\bibitem[al (2022)]{MINIMOSNT22}
{GTS Minimos-NT User Manual}. Global TCAD Solutions, Vienna, Austria,
  2022\relax
\mciteBstWouldAddEndPuncttrue
\mciteSetBstMidEndSepPunct{\mcitedefaultmidpunct}
{\mcitedefaultendpunct}{\mcitedefaultseppunct}\relax
\EndOfBibitem
\bibitem[{Alkauskas, A. and Yan, Q. and Van de Walle, C.G.}(2014)]{ALKAUSKAS14}
{Alkauskas, A. and Yan, Q. and Van de Walle, C.G.} {First-Principles Theory of
  Nonradiative Carrier Capture via Multiphonon Emission}. \emph{Phys. Rev. B}
  \textbf{2014}, \emph{90}, 075202\relax
\mciteBstWouldAddEndPuncttrue
\mciteSetBstMidEndSepPunct{\mcitedefaultmidpunct}
{\mcitedefaultendpunct}{\mcitedefaultseppunct}\relax
\EndOfBibitem
\bibitem[Illarionov \latin{et~al.}(2016)Illarionov, Waltl, Rzepa, Kim, Kim,
  Dodabalapur, Akinwande, and Grasser]{ILLARIONOV16B}
Illarionov,~Y.; Waltl,~M.; Rzepa,~G.; Kim,~J.-S.; Kim,~S.; Dodabalapur,~A.;
  Akinwande,~D.; Grasser,~T. {Long-Term Stability and Reliability of Black
  Phosphorus Field-Effect Transistors}. \emph{ACS Nano} \textbf{2016},
  \emph{10}, 9543--9549\relax
\mciteBstWouldAddEndPuncttrue
\mciteSetBstMidEndSepPunct{\mcitedefaultmidpunct}
{\mcitedefaultendpunct}{\mcitedefaultseppunct}\relax
\EndOfBibitem
\bibitem[Knobloch \latin{et~al.}(2018)Knobloch, Rzepa, Illarionov, Waltl,
  Schanovsky, Stampfer, Furchi, Mueller, and Grasser]{KNOBLOCH18}
Knobloch,~T.; Rzepa,~G.; Illarionov,~Y.; Waltl,~M.; Schanovsky,~F.;
  Stampfer,~B.; Furchi,~M.; Mueller,~T.; Grasser,~T. {A Physical Model for the
  Hysteresis in MoS$_2$ Transistors}. \emph{IEEE J. Electron Dev. Soc.}
  \textbf{2018}, \emph{6}, 972--978\relax
\mciteBstWouldAddEndPuncttrue
\mciteSetBstMidEndSepPunct{\mcitedefaultmidpunct}
{\mcitedefaultendpunct}{\mcitedefaultseppunct}\relax
\EndOfBibitem
\bibitem[Knobloch \latin{et~al.}(2023)Knobloch, Waldhoer, Davoudi, Karl,
  Khakbaz, Matzinger, Zhang, Smithe, Nazir, Liu, \latin{et~al.}
  others]{KNOBLOCH23}
Knobloch,~T \latin{et~al.} Modeling the Performance
  and Reliability of Two-Dimensional Semiconductor Transistors. IEEE
  International Electron Devices Meeting (IEDM). 2023; pp 1--4\relax
\mciteBstWouldAddEndPuncttrue
\mciteSetBstMidEndSepPunct{\mcitedefaultmidpunct}
{\mcitedefaultendpunct}{\mcitedefaultseppunct}\relax
\EndOfBibitem
\bibitem[Rasmussen and Thygesen(2015)Rasmussen, and Thygesen]{RASMUSSEN15}
Rasmussen,~F.; Thygesen,~K. {Computational 2D Materials Database: Electronic
  Structure of Transition-Metal Dichalcogenides and Oxides}. \emph{J.
  Phys. Chem. C} \textbf{2015}, \emph{119}, 13169--13183\relax
\mciteBstWouldAddEndPuncttrue
\mciteSetBstMidEndSepPunct{\mcitedefaultmidpunct}
{\mcitedefaultendpunct}{\mcitedefaultseppunct}\relax
\EndOfBibitem
\bibitem[Illarionov \latin{et~al.}(2016)Illarionov, Rzepa, Waltl, Knobloch,
  Grill, Furchi, Mueller, and Grasser]{ILLARIONOV16A}
Illarionov,~Y.; Rzepa,~G.; Waltl,~M.; Knobloch,~T.; Grill,~A.; Furchi,~M.;
  Mueller,~T.; Grasser,~T. {The Role of Charge Trapping in MoS$_2$/SiO$_2$ and
  MoS$_2$/hBN Field-Effect Transistors}. \emph{2D Mater.} \textbf{2016},
  \emph{3}, 035004\relax
\mciteBstWouldAddEndPuncttrue
\mciteSetBstMidEndSepPunct{\mcitedefaultmidpunct}
{\mcitedefaultendpunct}{\mcitedefaultseppunct}\relax
\EndOfBibitem
\bibitem[Illarionov \latin{et~al.}(2017)Illarionov, Knobloch, Waltl, Rzepa,
  Pospischil, Polyushkin, Furchi, Mueller, and Grasser]{IllARIONOV17A}
Illarionov,~Y.; Knobloch,~T.; Waltl,~M.; Rzepa,~G.; Pospischil,~A.;
  Polyushkin,~D.; Furchi,~M.; Mueller,~T.; Grasser,~T. {Energetic Mapping of
  Oxide Traps in MoS$_2$ Field-Effect Transistors}. \emph{2D Mater.}
  \textbf{2017}, \emph{4}, 025108\relax
\mciteBstWouldAddEndPuncttrue
\mciteSetBstMidEndSepPunct{\mcitedefaultmidpunct}
{\mcitedefaultendpunct}{\mcitedefaultseppunct}\relax
\EndOfBibitem
\end{mcitethebibliography}
